\documentclass[aps,prd,showpacs,twocolumn,nofootinbib,nobibnotes]{revtex4-1}

\RequirePackage{subfigure}
\usepackage{lipsum}
\usepackage{graphicx}
\usepackage{color}
\usepackage{enumerate}
\usepackage{amsmath}
\usepackage{bm}
\usepackage{comment}
\usepackage{diagbox}
\usepackage{amssymb}
\usepackage{stmaryrd}
\usepackage{multirow}
\usepackage[normalem]{ulem}
\usepackage{float}
\usepackage{longtable,booktabs}
\usepackage[dvipsnames,table,xcdraw]{xcolor}
\usepackage[normalem]{ulem}
\usepackage{physics}
\usepackage{url}
\usepackage{soul} 
\usepackage{verbatim}
\usepackage{siunitx}
\usepackage{booktabs,dcolumn}
\newcommand{\pislash}[0]{\pi\!\!\!/}

\definecolor{darkblue}{rgb}{0.1,0.2,0.6} \definecolor{darkred}{rgb}{0.8,0.1,0.2}
\usepackage[colorlinks,citecolor=darkblue,linkcolor=darkred,urlcolor=darkblue]{hyperref}
\usepackage[all]{hypcap} 

\usepackage[draft]{pdfcomment}

\def\beq{\begin{equation}}
\def\eeq{\end{equation}}	 

\def\hethree{$^{3}$He}
\def\hefour{$^{4}$He}

\def\hethreem{{}^{3}{\rm He}}
\def\hefourm{{}^{4}{\rm He}}

\begin{document}
\long\def\/*#1*/{}

\title{Finite-Volume Pionless Effective Field Theory for Few-Nucleon Systems with Differentiable Programming}

\author{Xiangkai Sun} 
\affiliation{
	Center for Theoretical Physics, 
	Massachusetts Institute of Technology, 
	Cambridge, MA 02139, USA}
\affiliation{The NSF AI Institute for Artificial Intelligence and Fundamental Interactions}

\author{William~Detmold} 
\affiliation{
	Center for Theoretical Physics, 
	Massachusetts Institute of Technology, 
	Cambridge, MA 02139, USA}
\affiliation{The NSF AI Institute for Artificial Intelligence and Fundamental Interactions}

\author{Di Luo} 
\affiliation{
	Center for Theoretical Physics, 
	Massachusetts Institute of Technology, 
	Cambridge, MA 02139, USA}
\affiliation{The NSF AI Institute for Artificial Intelligence and Fundamental Interactions}

\author{Phiala~E.~Shanahan} 
\affiliation{
	Center for Theoretical Physics, 
	Massachusetts Institute of Technology, 
	Cambridge, MA 02139, USA}
\affiliation{The NSF AI Institute for Artificial Intelligence and Fundamental Interactions}

\begin{abstract}

Finite-volume pionless effective field theory provides an efficient framework for the extrapolation of nuclear spectra and matrix elements calculated at finite volume in lattice QCD to infinite volume, and to nuclei with larger atomic number. In this work, it is demonstrated how this framework may be implemented via a set of correlated Gaussian wavefunctions optimised using differentiable programming and via solution of a generalised eigenvalue problem. This approach is shown to be significantly more efficient than a stochastic implementation of the variational method based on the same form of correlated Gaussian wavefunctions, yielding comparably accurate representations of the ground-state wavefunctions with an order of magnitude fewer terms. The efficiency of representation allows such calculations to be extended to larger systems than in previous work. The method is demonstrated through calculations of the binding energies of nuclei with atomic number $A\in\{2,3,4\}$ in finite volume, matched to lattice QCD calculations at quark masses corresponding to $m_\pi=806$~MeV, and infinite-volume effective field theory calculations of $A\in\{2,3,4,5,6\}$ systems based on this matching.
\end{abstract}
 
\maketitle

\section{Introduction}

A central goal of nuclear physics is to make predictions for the spectra and properties of nuclear systems based on the underlying degrees of freedom of the Standard Model, most pertinently quarks and gluons. 
Since nuclei and other relevant systems exhibit dynamics at energy scales where the interactions between quarks and gluons, governed in the Standard Model by the theory of Quantum Chromodynamics (QCD), are nonperturbative, this goal can be addressed directly only by numerical calculations in the framework of lattice QCD (LQCD). However, due to computational limitations, LQCD studies of nuclei have so far been restricted to systems with atomic number $A\le 4$, with unphysically large values of the quark masses. Moreover, to date only proof-of-principle LQCD calculations of nuclei have been performed \cite{Beane:2010hg,Inoue:2010es,Beane:2011iw,Beane:2012vq,Yamazaki:2012hi,Yamazaki:2015asa,Berkowitz:2015eaa,Orginos:2015aya,Francis:2018qch,Inoue:2015dmi,HALQCD:2012aa,Inoue:2018axd,Kawai:2017goq,Iritani:2018sra,Green:2021qol,Parreno:2021ovq}, in which systematic uncertainties such as those from the lattice discretisation are estimated but not fully quantified. 

While fully-controlled LQCD calculations of light nuclei will likely be achieved in the near future, the computational costs of such studies scale exponentially with $A$ in current approaches, and as such, the restriction to small nuclear systems is likely to persist until novel algorithms or other computational breakthroughs render calculations of larger nuclei tractable.
Pionless nuclear effective field theory (EFT) \cite{Kaplan:1996xu,Kaplan:1998tg,Kaplan:1998we,vanKolck:1998bw,Bedaque:1998kg,Chen:1999tn,Bedaque:2002yg,Bedaque:2002mn} provides a bridge between tractable LQCD calculations of light nuclei and the broader scope of low-energy nuclear phenomenology. 
In nuclear physics, it is apparent that there is a  hierarchy of interactions, in that two-body interactions are more important in governing nuclear structure and reactions than three-body interactions, which are in turn more important than four-body interactions, and so on. Consequently, LQCD calculations of $A\leq 4$ systems can be used to constrain the most relevant couplings in  nuclear EFT which can subsequently be used to make predictions for larger nuclear systems and for matrix elements which may not have been directly computed in LQCD. 
In addition, since the finite volume in which LQCD calculations are performed produces effects which are long-distance in nature, they can be captured in nuclear EFT calculations in appropriately matched finite volumes (finite volume nuclear EFT (FVEFT)). With the couplings of the EFT determined by this matching, the EFT provides a method to extract infinite volume physics from finite-volume LQCD spectra and matrix elements. 

Existing applications of FVEFT to the matching and extrapolation of LQCD results for nuclear spectra \cite{Eliyahu:2019nkz} and matrix elements \cite{Detmold:2021oro} have used the stochastic variational method (SVM) \cite{Varga:1995dm} with trial wavefunctions composed of shifted correlated Gaussian functions \cite{PhysRevA.87.063609}. Because of the stochastic nature of this approach, a large number of terms are required to approximate the ground state of each nuclear system. In this work, a new differentiable programming (DP) approach is introduced that implements an optimisation of the parameters defining each Gaussian term that is included in the trial wavefunction, as opposed to the stochastic selection of terms, resulting in much more efficient representations.
Further improvement through the combination of multiple sets of optimised trial states can be achieved through solution of a generalised eigenvalue problem (GEVP). The compactness of the resulting wavefunction representations makes it feasible to extend previous calculations to systems of larger $A$. In this work, FVEFT predictions are made for the $^4$He ground state as a function of volume, and the FVEFT matching of two and three-body interactions enables predictions for the infinite-volume energies of  $A\in\{5,6\}$ systems.

The following section outlines important aspects of nuclear FVEFT and the differential programming method used to determine optimal wavefunctions in the approach proposed here. Section \ref{sec:results} presents results of the optimisation procedure for  finite-volume systems with $A\in\{2,3,4\}$, and their matching to LQCD energy determinations. Infinite-volume
binding energies are also presented for $A\le6$. Section \ref{sec:conclusion} provides a summary and  outlook for this approach.

\section{Methodology}

\subsection{Hamiltonian for pionless effective field theory}

The low-energy interactions of nucleons are described in pionless EFT (EFT$_{\pislash}$) by the Lagrangian \cite{Kaplan:1996xu,Kaplan:1998tg,Kaplan:1998we,vanKolck:1998bw,Bedaque:1998kg,Chen:1999tn,Bedaque:2002yg,Bedaque:2002mn}
\begin{equation}
\begin{split}
    \mathcal{L} ={} & N^{\dagger}\left(i D_{0}+\frac{\mathbf{D}^{2}}{2 M_{N}}\right) N \\
    & -\frac{1}{2}\left[C_{0}\left(N^{\dagger} N\right)^{2}+C_{1}\left(N^{\dagger} \vec{\sigma} N\right)^{2} \right] \\
    & - \frac{D_0}{6}(N^\dagger N)^3  + \ldots.
\end{split}
\end{equation}
The first, second, and third lines present the leading-order single-nucleon kinetic operator expanded in the non-relativistic (NR) limit, the two-body interaction, and the three-body interaction, respectively (the latter is promoted to leading order to define a valid power-counting scheme). $N$ denotes the nucleon field, $M_N$ the nucleon mass, $\vec{\sigma}$ the vector of Pauli matrices acting in spin space of a given nucleon, and $\{C_0,C_1\}$ and $D_0$ denote the relevant two and three-body low-energy constants (LECs). A common alternate basis for the two-nucleon interactions yields related LECs
\begin{equation}
\label{eq:C01CSTrelation}
C_T=C_{0}-3 C_{1} \hspace{3mm}\text{and}\hspace{3mm} C_S=C_{0}+C_{1}.
\end{equation}

The corresponding $n$-particle non-relativistic Hamiltonian can be expressed as 
\begin{equation}
H=-\frac{1}{2 M_N} \sum_{i} \nabla_{i}^{2}+\sum_{i<j} V_{2}\left({\bf r}_{i j}\right)+\sum_{i<j<k} V_{3}\left({\bf r}_{i j}, {\bf r}_{j k}\right),
\label{eq:Hamiltonian}
\end{equation}
where the $n$ particles are labelled by indices $i,j,k\in \{1,\ldots, n\}$ and the Laplacian for particle $i$ is expressed as $\nabla^2_i$. $V_{2}\left({\bf r}_{i j}\right)$ and $V_{3}\left({\bf r}_{i j}, {\bf r}_{j k}\right)$ denote the two and three-particle potentials, which are regulated using Gaussian smearing, and are functions of the displacements between particles, defined for particles $i$ and $j$ as ${\bf r}_{ij} ={\bf r}_i - {\bf r}_j$, where ${\bf r}_i=(r_i^{(x)},r_i^{(y)},r_i^{(z)})$.
In particular,
\begin{align}
V_{2}\left({\bf r}_{i j}\right)=&\left(C_{0}+C_{1} \sigma^{(i)} \cdot \sigma^{(j)}\right) g_{\Lambda}\left({\bf r}_{i j}\right),\\
\intertext{and}
V_{3}\left({\bf r}_{i j}, {\bf r}_{j k}\right)={}&D_0 \sum_{\text{cyc}} g_{\Lambda}\left({\bf r}_{i j}\right) g_{\Lambda}\left({\bf r}_{j k}\right),
\end{align}
where $\sum_{\text{cyc}}$ denotes the sum over all cyclic permutations of $\{i,j,k\}$, and the Gaussian regulator in infinite spatial volume is defined as
\begin{align}\nonumber
g_{\Lambda}({\bf r})={}&\frac{\Lambda^{3}}{8 \pi^{3 / 2}} \exp \left(-\Lambda^{2} |{\bf r}|^{2} / 4\right)\\
={}&\frac{\Lambda^{3}}{8 \pi^{3 / 2}} \prod_{\alpha\in\{x,y,z\}} \exp \left(-\Lambda^{2} r^{(\alpha)2} / 4\right).
\label{eq:regulator}
\end{align}
The regulator parameter $\Lambda$ can be expressed in terms of a length-scale $r_0$ as $\Lambda=\sqrt{2}/r_0$. Physical quantities are independent of this cutoff \cite{Detmold:2021oro}. 

In a finite cubic spatial volume with side-length $L$, the regulator can be constructed to be periodic by summing $g_{\Lambda}({\bf r})$ over copies translated by multiples of $L$ in each spatial direction: 
\begin{align}
    g_\Lambda({\bf r},L) =&\frac{\Lambda^{3}}{8 \pi^{3 / 2}}  \prod_{\alpha\in\{x,y,z\}} \nonumber\\
    &{}\times\sum_{q^{(\alpha)}=-\infty}^\infty \exp \left(-\Lambda^{2} (r^{(\alpha)}-L q^{(\alpha)})^{2} / 4\right).
\label{eq:periodic_regulator}
\end{align}

\subsection{Variational method framework}

The variational method provides a systematically-improvable approach to bounding the ground (and excited) state energies of  quantum systems; given any wavefunction ansatz $\Psi_h\left({\bf x}\right)$ for a state $h$ defined over coordinates ${\bf x}$, the ground-state energy $E_h$ is bounded as 
\begin{equation}\label{eq:variationalGS}
E_{h} \leq {\cal E}[\Psi_h]=\frac{\int \Psi_{h}(\mathbf{x})^{*} H \Psi_{h}(\mathbf{x}) \,d \mathbf{x}}{\int \Psi_{h}(\mathbf{x})^{*} \Psi_{h}(\mathbf{x}) \,d \mathbf{x}}.
\end{equation}
A wavefunction ansatz that depends on some number of free parameters may be varied over those parameters to determine an
optimal bound within that ansatz class.

One approach to the variational method that has been successfully applied to the study of nuclear systems in a finite volume within the framework of pionless effective field theory is the stochastic variational method (SVM)~\cite{Varga:1995dm}. In this approach, a wavefunction is generated constructively through the iterative addition of stochastically-proposed terms, and a generalised eigenvalue problem is solved to optimise the linear combination of the proposed terms.
In particular, this approach has been applied in a finite volume in Refs.~\cite{Eliyahu:2019nkz,Detmold:2021oro} using a basis of correlated Gaussian terms. Here, the same wavefunction ansatz is considered, but is optimised using an alternative to the stochastic optimisation procedure that is based on differentiable programming (detailed in Sec.~\ref{subsec:diffprog}).

The Gaussian wavefunction ansatz used in this work is based on the approximation that the spatial and spin-isospin wavefunctions for nuclear states can be factorised, with spatial wavefunctions constructed as linear combinations of appropriately symmetrised Gaussians.\footnote{Although the factorisation of the spatial and spin wavefunctions is a crude approximation for larger nuclei, the goal of the present work is to explore the effectiveness of the differentiable programming approach in representing nuclear states in comparison to the stochastic variational method. As such, the same approximation is used as in Ref.~\cite{Detmold:2021oro}.} 
As also used in Refs.~\cite{Eliyahu:2019nkz,Detmold:2021oro}, a trial wavefunction of this form satisfying the periodic boundary conditions of a finite spatial volume can be expressed for some nucleus $h$ as  
\begin{equation}
    \Psi^{(N)}_h\left({\bf x}\right) =\sum_{j=1}^N c_j\Psi^{{\rm sym}}_{L}\left(A_j,B_j,{\bf d}_j; {\bf x}\right) |\chi_h\rangle,
    \label{eq:trialwf}
\end{equation}
where the sum runs over the $N$ terms included in the trial wavefunction, the $c_j$, $j\in\{1,\ldots,N\}$, are numerical coefficients, $|\chi_h\rangle$ denotes an appropriate normalised spin-flavour wavefunction for the $n$-body state $h$, ${\bf x}=({\bf r}_1,\ldots,{\bf r}_n)$ denotes the collected spatial coordinates of the $n$ nucleons, and the  $A_j$, $B_j$ and ${\bf d}_j$ denote collected parameters of the $j$th spatial wavefunction $\Psi^{{\rm sym}}_{L}$ included in the sum (whose dependence on $h$ is suppressed). To obtain an optimal representation of the wavefunction with a given number of terms, the values of the parameters $c_j$ and those encoded in $A_j$, $B_j$ and ${\bf d}_j$ are optimised as described further in Sec.~\ref{subsec:diffprog}.

Explicitly, the symmetrised spatial wavefunction $\Psi^{{\rm sym}}_{L}$ is constructed from Gaussian components for each Cartesian direction $\alpha$:
\begin{equation}
\begin{split}  
    \Psi_\infty^{(\alpha)}(A^{(\alpha)},B^{(\alpha)},{\bf d}^{(\alpha)}; {\bf x}^{(\alpha)})&=\exp \left[-\frac{1}{2} \mathbf{x}^{(\alpha)T} A^{(\alpha)} \mathbf{x}^{(\alpha)}\right.
   \\
    &\hspace*{-2.5cm}
  \left. -\frac{1}{2}(\mathbf{x}^{(\alpha)}-\mathbf{d}^{(\alpha)})^{T} B^{(\alpha)}(\mathbf{x}^{(\alpha)}-\mathbf{d}^{(\alpha)})\right]\,,
\end{split}
\end{equation}
where the $\alpha$th Cartesian components of the spatial coordinates of each particle are collected in the $n$-component vector ${\bf x}^{(\alpha)}$. The $n\times n$ matrices  $A^{(\alpha)}$ and $B^{(\alpha)}$ are symmetric, containing $n(n-1)/2$ real parameters, and diagonal, with $n$ real parameters, respectively, and ${\bf d}^{(\alpha)}$ is an $n$-component real-valued vector. This wavefunction can be made periodic in a cubic volume of finite spatial extent $L$ by implementing a sum over copies shifted in each Cartesian direction by integer multiples of $L$~\cite{PhysRevA.87.063609}:
\begin{align}    \nonumber
    \Psi^{(\alpha)}_{L}\left(A^{(\alpha)},B^{(\alpha)},{\bf d}^{(\alpha)}; {\bf x}^{(\alpha)}\right) &= 
    \\
 & \hspace*{-3cm}   \sum_{{\bf b}^{(\alpha)}} \Psi_\infty^{(\alpha)}(A^{(\alpha)},B^{(\alpha)},{\bf d}^{(\alpha)}; {\bf x}^{(\alpha)}- {\bf b}^{(\alpha)}L),
\end{align}
where ${\bf b}^{(\alpha)}$ is an $n$-component vector with components $b^{(\alpha)}_k\in \mathbb{Z}$. The finite-volume wavefunctions for each Cartesian direction $\alpha$ can be combined to define the complete three-dimensional finite-volume wavefunction
\begin{equation}
    \Psi_{L}\left(A,B,{\bf d}; {\bf x}\right) = \prod_{\alpha\in\{x,y,z\}} 
    \Psi^{(\alpha)}_{L}\left(A^{(\alpha)},B^{(\alpha)},{\bf d}^{(\alpha)}; {\bf x}^{(\alpha)}\right),
\end{equation}
where the parameters $A^{(\alpha)}$, $B^{(\alpha)}$, and ${\bf d}^{(\alpha)}$ for each Cartesian direction are combined into the quantities $A$, $B$ and ${\bf d}$.
Finally, a finite-volume wavefunction that is also symmetric under particle exchange can be constructed by explicitly symmetrising with respect to permutations of the rows and columns of $A^{(\alpha)}$ and $B^{(\alpha)}$ and of the rows of ${\bf d}^{(\alpha)}$, for all Cartesian components $\alpha$. Denoting the set of all such permutations as $\cal P$, a symmetric wavefunction ansatz can thus be expressed as
\begin{equation}
    \Psi^{{\rm sym}}_{L}\left(A,B,{\bf d}; {\bf x}\right) = \sum_{\cal P} \Psi_{L}\left(A_{\cal P},B_{\cal P},{\bf d}_{\cal P}; {\bf x}\right),
    \label{eq:symwf}
\end{equation}
where $A_{\cal P}$, $B_{\cal P}$ and ${\bf d}_{\cal P}$ are the permuted forms of the relevant matrices and vectors.

A particular advantage of this class of trial wavefunctions is that the integrals needed to compute the normalisation and Hamiltonian matrix elements that appear in the ground-state energy bound of Eq.~\eqref{eq:variationalGS} can be performed analytically, as detailed in Ref.~\cite{Detmold:2021oro}. As also discussed in Ref.~\cite{Detmold:2021oro}, these Gaussian-based wavefunctions are able to represent finite-volume ``scattering states'', i.e., eigenstates above the two-particle threshold, for $N=2$ systems, and the method does not rely on deeply-bound infinite volume states. The only restrictions on its applicability are that states that are integrated out of the pionless EFT, such as those involving pions,  $\Delta$-resonances and particle--anti-particle excitations,  are not representable. These restrictions are similar to those in the L\"uscher quantisation condition approach \cite{Luscher:1986pf} where the partial-wave expansion of scattering amplitudes must be truncated and the presence of inelastic thresholds limits applicability.

\subsection{Variational optimisation by differentiable programming}
\label{subsec:diffprog}

To achieve effective bounds on the ground-state energies $E_h$ of various nuclear systems, $h$, the trial wavefunction $\Psi^{(N)}_h\left({\bf x}\right)$ defined in Eq.~\eqref{eq:trialwf} is optimised using a differentiable programming approach combined with solution of a GEVP. 
Differentiable programming is a programming paradigm in which  the computational flow of a program can be explicitly differentiated with respect to its parameters, thereby allowing gradient-descent optimisation of those parameters.
The approach is widely used as a backbone of machine learning tools~\cite{JMLR:v18:17-468} and has been applied to variational problems in quantum many-body physics~\cite{2021arXiv211011678K,2019PhRvX...9c1041L,arrazola2021differentiable,2018adhf} and quantum technology \cite{2021DP_transport,2020DP_qcontrol,khait2021optimal}.

\begin{figure*}[!t]
    \centering
    \subfigure[\ Differentiable programming (DP) block]{\includegraphics[width=\linewidth]{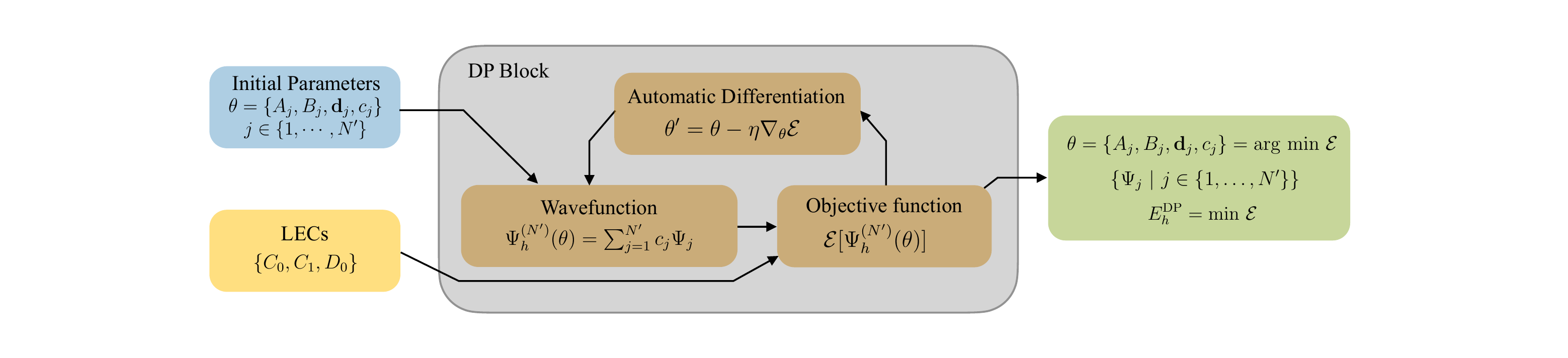}}\\ 
    \subfigure[\ Generalised eigenvalue problem (GEVP) block.]{\includegraphics[width=\linewidth]{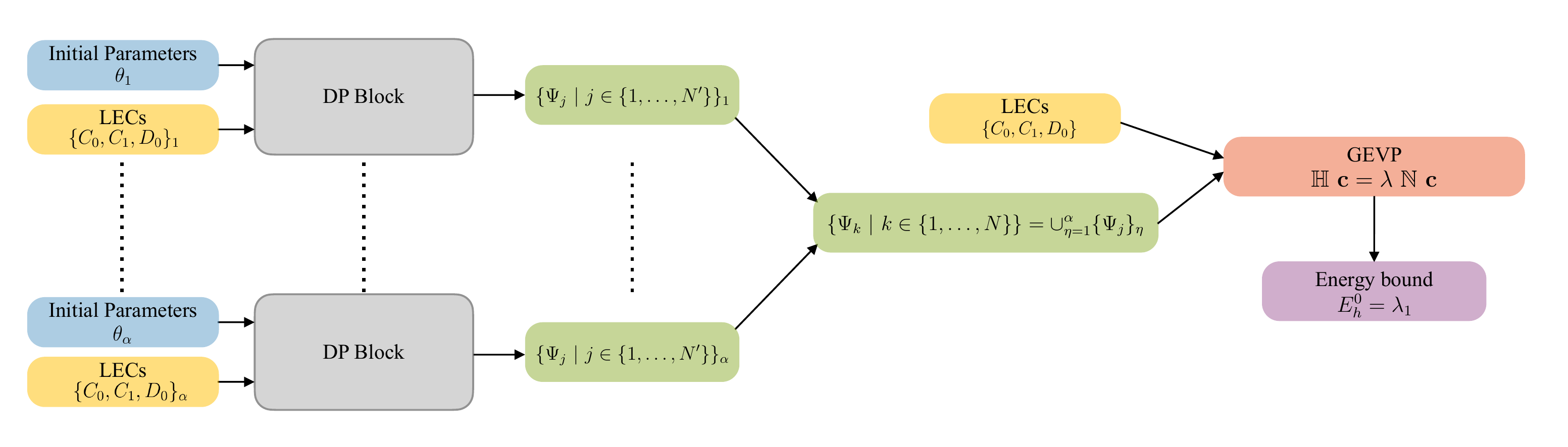}}
    \caption{Diagrammatic representation of the wavefunction optimisation procedure used in this work. (a) Differentiable programming block: for fixed LECs and a random initialisation, automatic differentiation (defined in the figure, where $\eta$ denotes the self-adaptive learning rate) is used to optimise the parameters of an $N'$-term Gaussian wavefunction ansatz. (b) GEVP block: basis elements obtained from  differentiable programming blocks constructed with different initialisations and/or LECs are combined to form a larger basis; the GEVP as defined in Eq.~\eqref{eq:GEVP} 
    is solved to determine the optimal energy bound $E_h^0$, the smallest eigenvalue, for a given set of LECs.
    }
    \label{fig:CG}
\end{figure*}

In the current application, the differentiable programming approach is applied to ground-state energy minimisation through a two-stage procedure:
\begin{enumerate}
    \item Differentiable program (DP) block: optimisation of $N'$-element wavefunctions with fixed LECs:
    \begin{itemize}
    \item Values of the LECs $C_0$, $C_1$, and (for $n\ge 3$ body states) $D_0$ are chosen;
    \item The free parameters $c_j$, and those encoded in $A_j$, $B_j$ and ${\bf d}_j$, for $j\in\{1,\ldots N'\}$ for a $N'$-term Gaussian wavefunction $\Psi^{(N')}_h({\bf x})$ are initialised randomly (details of the choice of initialisation for the numerical study detailed in Sec.~\ref{sec:results} are provided in Appendix~\ref{app:op});
    \item The gradient of the ground-state energy bound provided by the trial wavefunction with respect to the free parameters is computed, and the minimum is approached via gradient descent to optimise $c_j$ and the parameters in $A_j$, $B_j$ and ${\bf d}_j$ (details of the gradient computation and descent method are presented in Appendices~\ref{sec:appDP} and \ref{app:op}).
    \end{itemize}
    \item GEVP block: $N$-element wavefunction construction:
    \begin{itemize}
        \item A set of $\alpha$ $N'$-element wavefunctions, possibly optimised using different LECs $C_0, C_1, D_0$, and different initialisations, but defined for the same quantum numbers (number of particles and spin-flavour structure) and finite spatial extent, $L$, are constructed through $\alpha$ independent DP blocks;
        \item For a fixed choice of LECs, the linear combination of the optimised Gaussian wavefunction components (i.e., the $N=\alpha N'$ Gaussians with each optimised choice of $A$, $B$, $\bf d$) is optimised by solving the GEVP 
        \begin{equation} \label{eq:GEVP}
            \mathbb{H}\; {\bf c}=\lambda \mathbb{N}\;{\bf c},
        \end{equation}
        for the eigenvalues $\lambda_1\leq \lambda_2\leq\ldots\leq \lambda_N$ and eigenvectors ${\bf c}=(c_1,\ldots,c_N)^T$ which contain the coefficients in Eq.~\eqref{eq:trialwf}. The matrices $\mathbb{H}$ and $\mathbb{N}$ have matrix elements
     \begin{align}\label{eq:HNint}
        [\mathbb{N}]_{ij}\equiv& \int \Psi_i({\bf x})^\ast 
        \Psi_j({\bf x}) d{\bf x},\\\label{eq:Hint}
         [\mathbb{H}]_{ij}\equiv&\int \Psi_i({\bf x})^\ast 
        \langle \chi_h| H  |\chi_h\rangle \Psi_j({\bf x}) d{\bf x}, 
    \end{align}  
        using the compressed notation $\Psi_j({\bf x}) \equiv \Psi^{{\rm sym}}_{L}\left(A_j,B_j,{\bf d}_j; {\bf x}\right)$.
        
    \item The lowest eigenvalue, $\lambda_1$, of the GEVP solved for a given set of LECs, $C_0, C_1, D_0$, corresponds to an upper bound $E_h^0$ on the ground state energy $E_h$ for the given system with quantum numbers defined by $h$.
        
    \end{itemize}
\end{enumerate}

This approach, illustrated graphically in Fig.~\ref{fig:CG}, has several advantages. First, the use of direct optimisation as opposed to stochastic selection of Gaussian basis elements enables wavefunction representations of comparable quality to be obtained with far fewer terms, as demonstrated in numerical experiments detailed in Sec.~\ref{subsec:illustration}. 
Second, this particular (sequential) optimisation approach enables the efficient construction of $N$-term wavefunctions by combining the Gaussian basis elements obtained by optimising systems with fewer terms; this is computationally efficient since the cost of directly optimising an $N$-term wavefunction grows quadratically with $N$.\footnote{In particular, the cost of optimisation of an $n$-body state with $N$ terms scales as $O(N^2n!n^3)$. While this complete optimisation would in principle outperform the sequential approach used here for a fixed number of terms, the sequential approach is superior for a fixed computational budget, scaling as $O(\alpha {N'}^2n!n^3)$.} Simultaneously, by combining wavefunctions optimised for different choices of LECs, this approach enables the construction of a combined set of Gaussian terms that can provide efficient wavefunction representations across a range of values of the LECs. 
With such a basis defined, constraining the LECs to match the FVEFT to LQCD calculations of nuclear states in the same finite volume is straightforward; computing the energy bound as a function of the LECs simply amounts to repeating the GEVP for choices of the LECs within a range of interest (involving no additional differentiable programming optimisation).

\section{Results}
\label{sec:results}

The differentiable programming approach described in Sec.~\ref{subsec:diffprog} is applied to the determination of ground-state energies of $A\in\{2,3,4,5,6\}$ nuclear systems, via optimisation of spatial nuclear wavefunctions with the relevant LECs tuned to match the results of LQCD calculations for $A\in \{2,3\}$.
As was previously investigated in the SVM in Refs.~\cite{Eliyahu:2019nkz,Detmold:2021oro}, the differentiable programming method can be used to extrapolate existing LQCD results for light nuclei to infinite volume. However, the more efficient representation provided by the DP wavefunctions also allows extrapolation of the LQCD results to systems with larger $A$.

\subsection{Illustration of differentiable programming optimisation}
\label{subsec:illustration}

This section provides a numerical illustration of the differentiable programming approach of Sec.~\ref{subsec:diffprog}.  The following examples demonstrate each step of the method, while discussion of several more technical aspects of the approach such as the initialisation of the free parameters, the schedule of optimisation (`training'), and the convergence criteria used in the applications in the following sections, are left to Appendix~\ref{app:op}.

\begin{figure}[!th]
    \centering
    	\subfigure[]{\includegraphics[width=\linewidth]{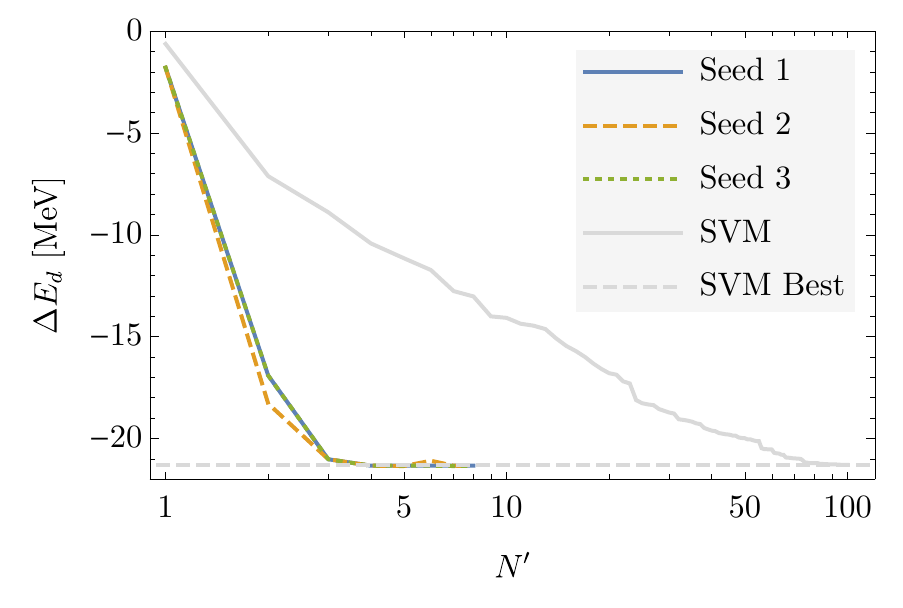}}\\
    	\subfigure[]{\includegraphics[width=\linewidth]{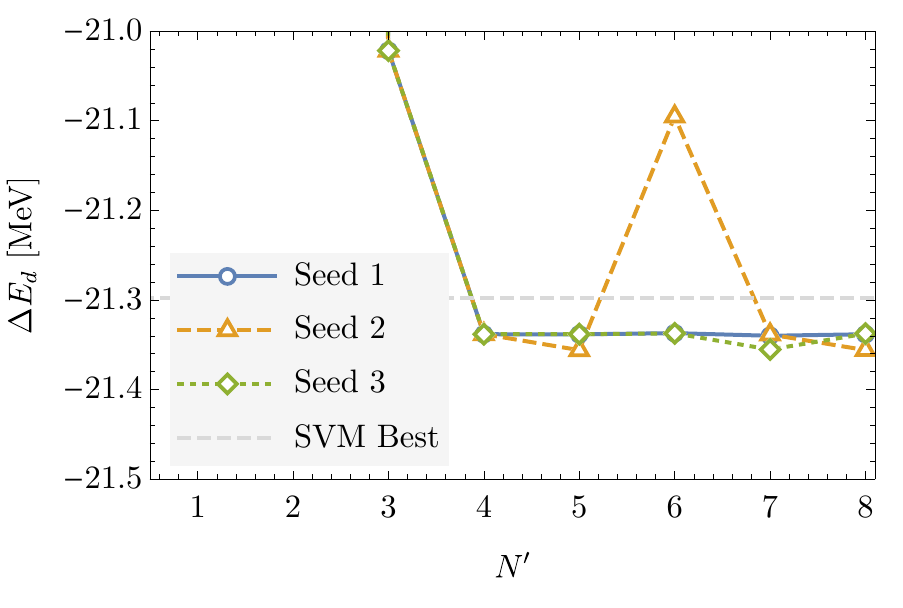}}
    \caption{The bounds on the binding energy of the deuteron obtained from  DP blocks as a function of the number of Gaussian functions, $N'$, included in the optimisation. Three different random initialisations are used (corresponding to the solid blue, dashed orange, and dotted green curves) for fixed values of the relevant two-body LEC $C_S=-132~\textrm{MeV}\cdot~\textrm{fm}^3$, and for $L=4.5$~fm. In the upper panel (a), the results are compared with the SVM results of Ref.~\cite{Detmold:2021oro} evaluated at the same $C_S$ and $L$, shown as a function of the number of Gaussian functions (solid grey line). The lower panel (b) shows the DP results with a different scale.
    The dashed horizontal line in both panels shows the best result obtained with the SVM method with $N'=100$.
    }
    \label{fig:convLog1}
\end{figure}

As discussed in Sec.~\ref{subsec:diffprog}, the differentiable programming optimisation procedure proceeds via DP blocks and GEVP blocks. The DP block step yields an optimised $N'$-term wavefunction at fixed LECs; Fig.~\ref{fig:convLog1} provides an example of the convergence of this optimisation with $N'$. In particular, the figure illustrates the bound on the binding energy of the $A=2$ deuteron ($h=d$) system achieved through a DP block optimisation (i.e., $\Delta E_d\equiv E_d - 2 E_p$, with $E_d=E_d^\text{DP}$) for a fixed choice of the relevant LEC $C_S=-132 ~\textrm{MeV}\cdot \textrm{fm}^3$, and spatial volume $L=4.5$~fm (these parameters are approximately in the centre of the ranges that are used in the application of the method in the following sections). Clearly, an improved bound on the binding energy is achieved with increasing $N'$, although this improvement need not be monotonic since the optimisation is performed from a new initialisation for each $N'$.\footnote{An alternate approach in which an optimised $N'-M$ term wavefunction is used to build an $N'$ term wavefunction by only optimising the parameters associated with the $M$ new Gaussian functions (through DP) and the linear coefficients (through GEVP) could also be applied, and would be monotonic by construction.} Different initialisation seeds typically yield consistent results for $N'\agt 4$. The figure also shows the result of the SVM optimisation method from Ref.~\cite{Detmold:2021oro}, demonstrating that the DP optimisation procedure provides a far more efficient description of the ground state in terms of the number of parameters that are required; for most initialisation seeds, the DP wavefunctions with $N'\agt 4$ outperform the $N'=100$ term wavefunction of Ref.~\cite{Detmold:2021oro}. 

\begin{figure}[!t]
    \centering
    \includegraphics[width=\linewidth]{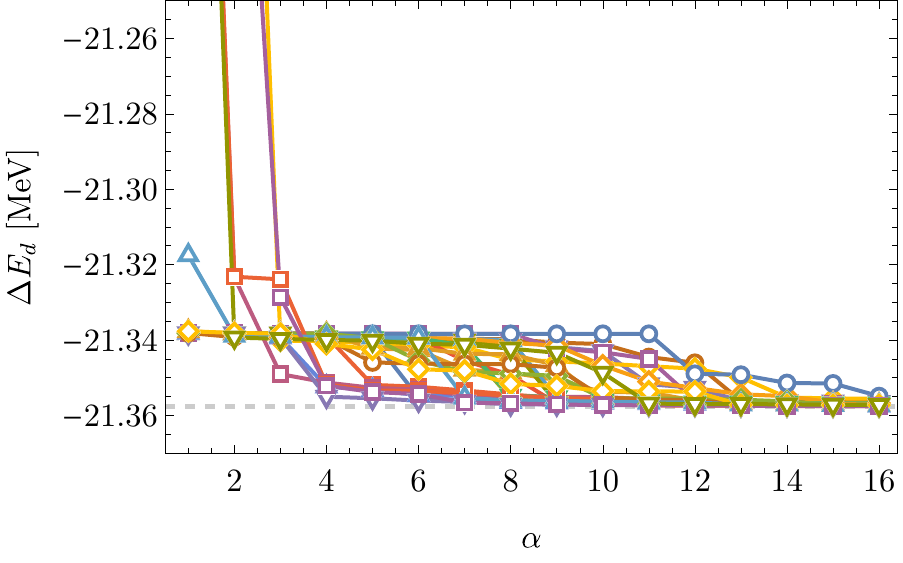}
    \caption{Bounds on the binding energy of the deuteron at $C_S=-132 ~\textrm{MeV}\cdot \textrm{fm}^3$ and $L=4.5~\textrm{fm}$ as a function of the number of groups of $N'=4$ Gaussian wavefunctions that are combined through a GEVP block. Each of the 25 curves shows bounds obtained using a different random sampling of groups of 4 Gaussians from a total of 36 groups, each optimised from a different random initialisation. The dashed grey line shows the tightest bound on the binding energy achieved by any of the optimisations. }
    \label{fig:convG}
\end{figure}

The second step of optimisation  combines $\alpha$ sets of $N'$ Gaussian functions determined in independent DP blocks through a GEVP block to determine the optimal linear combination of all $N=\alpha N'$ Gaussian functions. As $\alpha$ increases, the bound on the ground-state energy of the system necessarily improves. Figure~\ref{fig:convG} shows the binding energy of the deuteron with $C_S=-132 ~\textrm{MeV}\cdot \textrm{fm}^3$ and $L=4.5$~fm obtained from GEVP-optimised combinations of $\alpha\in\{1,\ldots,16\}$ sets of $N'=4$ Gaussian functions. Results are shown for 25 different initialisations. For $\alpha\ge 8$, all seeds yield values within 0.1\% of the minimum energy achieved with $\alpha=16$ groups of Gaussians (with any seed), and at $\alpha=8$, more than half of the seeds yield results within 0.05\% of that minimum. 

\begin{figure}[!t]
    \centering
    \includegraphics[width=1\linewidth]{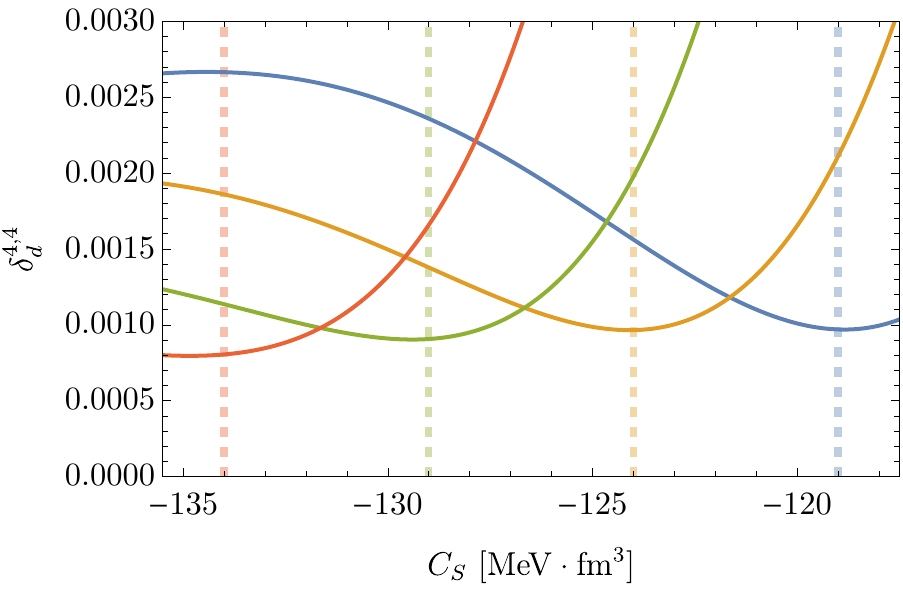}
    \caption{Fractional difference between the binding energy of the deuteron at $L=4.5~\textrm{fm}$ computed via GEVP from $\alpha=4$ DP blocks optimised at $C_S\in\{-134,-129,-124,-119\}~\mathrm{MeV\cdot fm^3}$ with $N'=4$, and the results of GEVP using each block separately. Each curve corresponds to $\delta_d^{4,4}$ (Eq.~\eqref{eq:deltah}) computed based on the DP block optimised at the value of $C_S$ indicated by the colour-matched vertical dotted line).}
    \label{fig:diff}
\end{figure}

While Fig.~\ref{fig:convG} illustrates the results of a GEVP block combining wavefunctions from several DP blocks optimised at the same set of LECs, a potentially useful alternative is to combine sets of Gaussian functions from DP blocks optimised at different choices of LECs. This produces a set of functions that should be better able to represent the eigenstates of the Hamiltonian across a range of values of the LECs, allowing energy bounds to be evaluated as a function of the LECs without additional DP optimisations.
For $\alpha$ sets of $N'$ Gaussian wavefunctions optimised in DP blocks, the quantity
\begin{equation} 
\label{eq:deltah}
    \delta_h^{\alpha,N'} = \frac{\Delta E^{[N']}_h-\Delta E^{[\alpha \times N']}_h }{\big|\Delta E^{[\alpha \times N']}_h\big|}
\end{equation}
can be defined to quantify the relative improvement of the combined  $\alpha\times N'$-term wavefunction (yielding a bound $\Delta E^{[\alpha \times N']}_h$ on the binding energy) over the $N'$ term wavefunction (yielding the bound $\Delta E^{[N']}_h$) at a given LEC value. Fig.~\ref{fig:diff} shows this quantity for the deuteron at $L=4.5~\textrm{fm}$, where sets of $N'=4$ Gaussian functions optimised at four choices of $C_S$ are combined in a GEVP block. By construction, GEVP-optimisation of the superset of 16 Gaussians provides a tighter bound on the binding energy across all LECs in the relevant range, improving the bound by $\lesssim 0.1$\% even at the LEC values where the individual DP blocks were optimised.

\subsection{Finite-Volume Calculation of Two-body and Three-body Systems}
\label{sec:3b}

In order to determine the two and three-body LECs in the FVEFT Hamiltonian,  $C_0$ and $C_1$ (or equivalently, $C_S$ and $C_T$) and $D_0$, wavefunction optimisations are performed using the approach of Sec.~\ref{subsec:diffprog} for the deuteron, dineutron, and \hethree\ systems in each of three spatial volumes where LQCD calculations have been performed~\cite{Beane:2012vq}. In this work, a single EFT cutoff corresponding to $r_0=0.2$ fm is used, as Refs.~\cite{Eliyahu:2019nkz,Detmold:2021oro} have previously demonstrated the cutoff-independence of the  ground-state energies.

\begin{figure}[!t]
    \centering
    \includegraphics[scale=0.95]{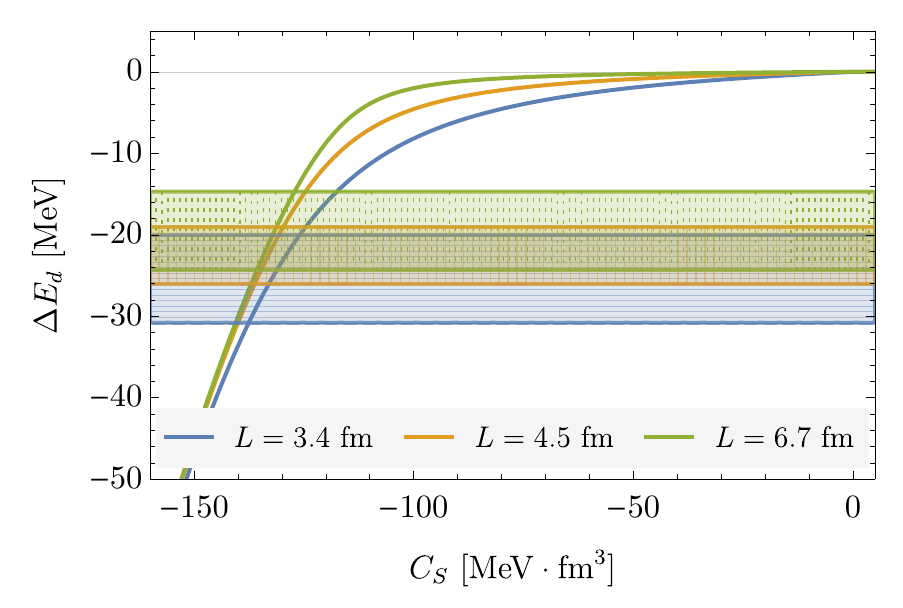}
    \caption{Binding energy of the deuteron as a function of the two-body LEC $C_S$. Each curve is obtained by solving the GEVP with various $C_S$ values using a 32-dimensional set of Gaussian wavefunctions optimised for each volume. Each set is composed of $\alpha=8$ sets of $N'=4$ DP blocks, with two blocks optimised from different initalisations for each $C_S\in\{-134,-129,-124,-119\}~\textrm{MeV}\cdot \textrm{fm}^3$. The horizontal bands show the binding energies determined in each volume in the LQCD calculations of Ref.~\cite{Beane:2012vq}.
    The intersection of each curve with the horizontal band of the same colour constrains the allowed values of $C_S$ through $\chi^2$-minimisation.
    }
    \label{fig:vsc2}
\end{figure}

Fig.~\ref{fig:vsc2} shows the binding energy of the deuteron as a function of the LEC $C_S$ for $L\in\{3.4,4.5,6.7\}~\textrm{fm}$.
The dependence of the binding energy in each volume on $C_S$ is obtained by solving the GEVP using a 32-dimensional basis of Gaussians, with $\alpha=8$ sets of $N'=4$ Gaussians, two optimised from different initialisations at each $C_S\in\{-134,-129,-124,-119\}~\textrm{MeV}\cdot \textrm{fm}^3$.
These choices of $N'$ and $\alpha$ achieve a balance between representational flexibility and computational cost and are motivated by the observations illustrated in Sec.~\ref{subsec:illustration}. In particular, taking $N'>4$ typically does not improve the bound achieved by a single DP block, and the combination of $\alpha=8$ DP blocks with $N'=4$ yields results within a fraction of a percent of the best result obtained by continuing to increase the number of blocks included; this difference is negligible in comparison with the uncertainties of the LQCD results used to match the LECs. The optimisation procedure and convergence criteria are detailed in Appendix~\ref{app:op}.
The same optimisations give the dependence of the dineutron binding energy on the LEC $C_T$; $C_S$ and $C_T$ can thus be obtained by $\chi^2$-minimisation of the difference between the optimised binding energies and the LQCD results of Ref.~\cite{Beane:2012vq} for the deuteron and dineutron, respectively. Fit results are shown in Table~\ref{tab:coeff} and are consistent within uncertainties with those obtained in Ref.~\cite{Detmold:2021oro} using the SVM approach to wavefunction optimisation, matched to the same LQCD results.  

\begin{figure}[tbp]
    \centering
    \includegraphics[scale=0.95]{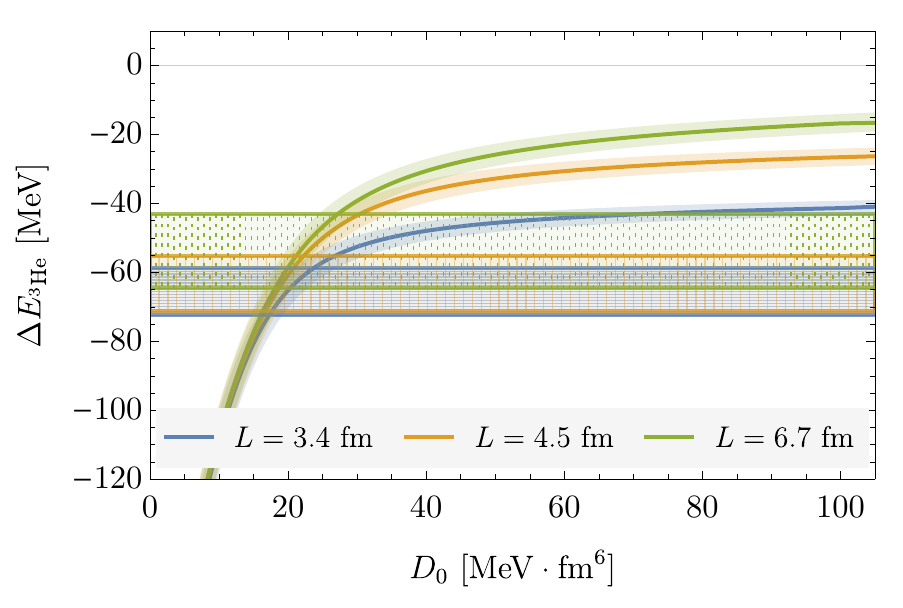}
    \caption{Binding energy of \hethree\ as a function of the three-body LEC $D_0$, with the relevant two-body coupling fixed to $C_{\hethreem}=C_0-C_1=129(2)\ \textrm{MeV}\cdot \textrm{fm}^3$. The curves are obtained by solving the GEVP for a 32-dimensional set of Gaussian wavefuntions optimised for each volume. Each set is composed of $\alpha=8$ $N'=4$ DP blocks, with two blocks optimised from different initalisations for each $D_0\in\{16.9,18.1,19.4,20.6\}~\textrm{MeV}\cdot \textrm{fm}^6$. The shaded region for each curve is propagated from the uncertainty in $C_{\hethreem}$. The horizontal bands show the binding energies determined in each volume in the LQCD calculations of Ref.~\cite{Beane:2012vq}.
    } 
    \label{fig:vsc3}
\end{figure}

Having determined the two-body couplings, the analogous procedure can be applied to determine the three-body interaction coefficient, $D_0$.  The GEVP is solved using a 32-dimensional basis of three-body Gaussians with $\alpha=8$ sets of $N'=4$ Gaussians, two optimised from different initialisations at each $D_0\in\{17.8,18.8,19.7,20.6\}~\textrm{MeV}\cdot \textrm{fm}^6$, with the optimised value of $C_{\hethreem}=C_0-C_1$ fixed.
The three-body binding energy is shown as a function of $D_0$ in Fig.~\ref{fig:vsc3} and the value of the coupling determined by $\chi^2$-minimisation of results at all three volumes is presented in Table.~\ref{tab:coeff}. 

\newcolumntype{F}{D{.}{.}{4}}
\begin{table}[tbp]
    \centering
        \begin{tabular}{cFF}    
    \hline \hline
    \toprule
 & \multicolumn{1}{c}{\text{Ref.~\cite{Detmold:2021oro}\ (SVM)}} & \multicolumn{1}{c}{\text{This\ work~(DP)}} \\            
        \hline
    \midrule
        $C_0$  & -131(2) & -131.0(21) \\
        $C_1$   & -2(1) & - 1.7(8) \\
        \hline
        $C_S$   & -133(2) & -132.7(27) \\
        $C_T$    & -126(2) & -125.8(20) \\
        \hline
        $D_0$  & 17(2)  & 20.0(24) \\
        \hline \hline
    \end{tabular}
    \caption{LECs in the EFT Lagrangian for a cutoff $r_0=0.2$~fm. $C_{S,T,0,1}$ and $D_0$ are quoted in units of $\text{MeV}\cdot \text{fm}^3$ and  $\text{MeV}\cdot \text{fm}^6$, respectively. The second column provides a comparison with the results obtained in Ref.~\cite{Detmold:2021oro} using the SVM, while the third column presents the results of this work obtained via the DP approach. }
    \label{tab:coeff}
\end{table}

\subsection{Finite-Volume Calculation of \hefour\ }
\label{sec:3c}

With all of the leading-order couplings in the EFT Lagrangian determined, the DP approach can be used to compute the ground-state energies of larger systems. In particular, an upper bound on the ground-state energy of \hefour\ is computed; with the LECs fixed, $\alpha=8$ $N'=4$ DP blocks optimised from different initialisations are combined via a GEVP block in each of the three spatial volumes in which LQCD calculations have been performed. After optimisation, the uncertainties in the two and three-body LECs are propagated into the estimate of the \hefour\ binding energy by solving the GEVP for each optimised set of Gaussian wavefunctions with the LECs varied within their uncertainty ranges. Table.~\ref{tab:he4} shows a comparison between the resulting \hefour\ binding energy and LQCD results for the  binding in each finite volume. Clearly, the EFT with fixed coefficients produces estimates of the \hefour\ binding energy that are consistent with the LQCD calculations. Having verified the consistency, an alternative strategy is to use the full set of $h\in\{d,nn,$\hethree,\hefour\} binding energies from LQCD to further constrain the two and three-body LECs. However, given the large uncertainties on the LQCD determinations of the \hefour\ energy, $\chi^2$-optimisation leads to values of the two and three-body LECs that are identical to those determined from the two and three-body systems alone.

In principle, the DP method can be used to determine finite-volume energies of still larger nuclei. However, there are currently no LQCD results to compare to for larger systems, and the scaling of the approach with $A$ at finite volume is sufficiently poor that such calculations are numerically demanding. Instead, it is natural to consider the infinite-volume binding energies of larger nuclei, as detailed in the next subsection.

\begin{table}[!t]
    \centering
    \begin{tabular}{ccc}
\hline \hline $L\ \mathrm{[fm]}$& LQCD~\cite{Beane:2012vq} [MeV] & This work [MeV] \\
\hline $3.4$ & $115(23)$ & $114(13)$  \\
$4.5$ & $107(25)$ & $109(15)$ \\
$6.7$ & $107(24)$ & $108(15)$ \\
\hline \hline
\end{tabular}
    \caption{Finite-volume binding energy for $^4$He obtained for three different  volumes as described in the text, with LECs $C_0$, $C_1$ and $D_0$ determined by matching to the two and three-body finite-volume LQCD results of Ref.~\cite{Beane:2012vq}. The second column lists the LQCD results for $\Delta E_\textrm{\hefour}$ computed in the same reference for comparison.}
    \label{tab:he4}
\end{table}

\subsection{Volume Dependence and Infinite-Volume Calculation}
\label{sec:3d}

\begin{figure}[tbp]
    \centering
    \includegraphics[width=\columnwidth]{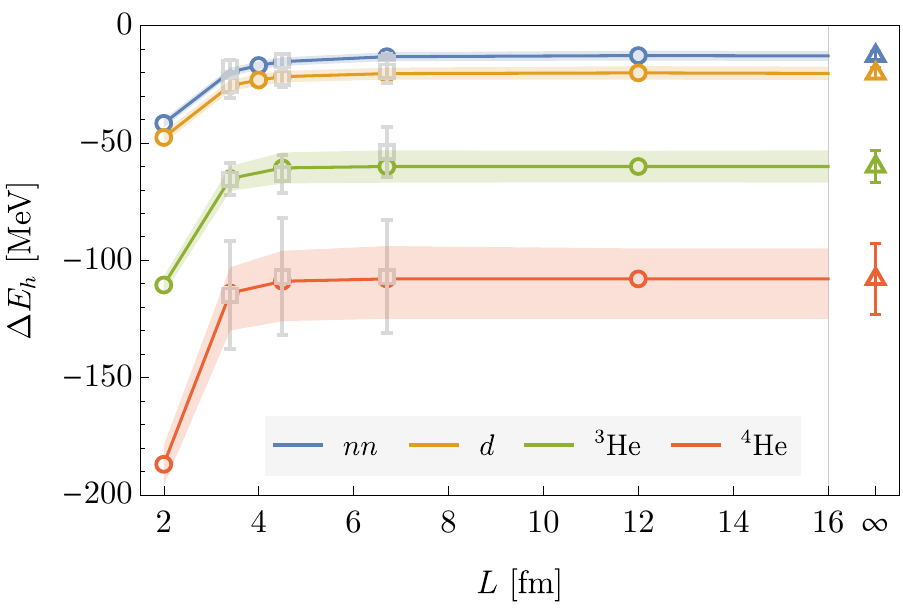}
    \caption{Binding energies of $h\in\{nn,d,\hethreem,\hefourm\}$ states in different volumes (open circles) and in infinite volume (open triangles). For each state at each of the volumes indicated by the open circles, $\alpha=8$ sets of $N'=4$ Gaussian wavefunctions are generated and optimised from different initialisations at the best fit values of the LECs $C_{0,1}$ and $D_0$ (the coloured lines are linear interpolations to guide the eye). The bands result from propagation of the uncertainties in the LECs. 
    The gray squares show the results of the LQCD calculations of Ref.~\cite{Beane:2012vq}. 
    }
    \label{fig:2pL}
\end{figure}

Having performed the finite-volume matching to determine the LECs of the EFT, the DP approach can be used to study the volume-dependence of the binding energies for $h\in\{d,nn,$\hethree,\hefour\}, as well as to determine the infinite-volume binding energies of these and other nuclear states.
Fig.~\ref{fig:2pL} shows the volume-dependence of the binding energies of the four systems, with optimisations based on $\alpha=8$ sets of $N'=4$ Gaussian wavefunctions performed for $L\in\{2,3.4,4,4.5,6.7,12 \}$~fm, and also at infinite volume. The infinite-volume results are compiled in Table \ref{tab:inf}.

\begin{table*}[!t]
    \centering
    \begin{tabular}{ccccccccc}
\hline \hline System & $I$ & $S$ & $J$ & Ref.~\cite{Beane:2012vq} & Ref.~\cite{Eliyahu:2019nkz} & Ref.~\cite{Detmold:2021oro} & Ref.~\cite{Barnea:2013uqa} & This work \\
\hline $nn$ & 1 & 0 & 0 & $16(4)$ & $14(2)$ & $12(2)$ & $16(2)$  & $12(2)$ \\
$d$ & 0 & 0 & 1 & $20(5)$ & $20(2)$ & $20(3)$ &$20(5)$& $20(3)$\\
${ }^{3} \mathrm{He}$ & 1/2 & 0 & 1/2 & $54(11)$ & $58(5)$ &  $60(7)$ & $54(11)$& $60(7)$\\
${ }^{4} \mathrm{He}$ & 0 & 0 & 0 & $107(24)$ & $113(10)$ & -- & $89(36)$& $108(15)$\\
${ }^{5}_{\Lambda} \mathrm{He}$ & 0 & -1 & 1/2 & -- & --  & & $98(39)$ & $162(24)$\\
${ }^{6}_{\Lambda\Lambda} \mathrm{He}$ & 0 & -2 & 0 &-- & -- & --& $122(50)$ & $215(32)$\\
\hline \hline
\end{tabular}
    \caption{Infinite-volume binding energies for various nuclear systems obtained in this work, compared with the  extrapolations in Refs.~\cite{Beane:2012vq} (LQCD), \cite{Eliyahu:2019nkz} (SVM), and \cite{Detmold:2021oro} (SVM), as well as the infinite-volume EFT calculations of Ref.~\cite{Barnea:2013uqa}. The isospin, $I$, strangeness, $S$, and spin, $J$, of each state is also listed. }
    \label{tab:inf}
\end{table*}
\begin{figure}[!tbp]
    \centering
    \includegraphics[width=\columnwidth]{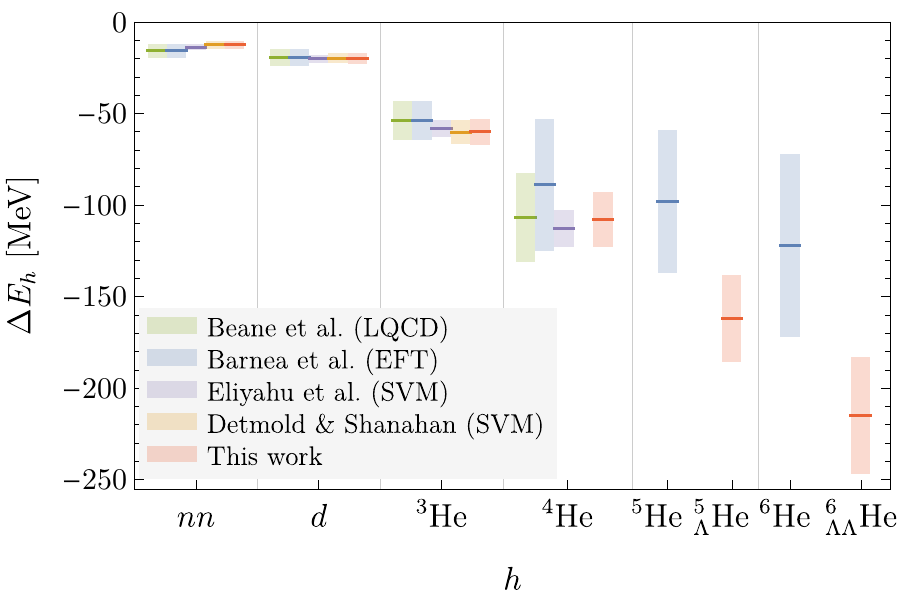}
    \caption{Comparisons of extrapolated infinite-volume binding energies obtained in different approaches: LQCD from Ref.~\cite{Beane:2012vq}, infinite-volume EFT matching from Ref.~\cite{Barnea:2013uqa}, and finite-volume EFT matching from Refs.~\cite{Eliyahu:2019nkz,Detmold:2021oro} and the current work.
    In each case, the EFT matching is performed to the same LQCD calculations of $h\in\{nn,d,$\hethree\} systems. 
    }
    \label{fig:infvol}
\end{figure}

Since the DP approach is more efficiently able to represent ground-state wavefunctions than the SVM method, it is feasible  to extend calculations to larger nuclei in infinite volume. In particular, calculations have been performed for $^{5}_{\Lambda}{\rm He}$ ($J=1/2$) and  $^{6}_{\Lambda\Lambda}{\rm He}$ ($J=0$) in which the spin-flavour structure is such that the simplest configuration has spatial and spin-flavour wavefunctions factorising as in Eq.~\eqref{eq:trialwf}.
Note that at the $SU(3)_f$-symmetric quark masses used in the LQCD calculations of Ref.~\cite{Beane:2012vq} to which the FVEFT calculation is matched, the $\Lambda$ baryon is degenerate with the proton and neutron, but is not Pauli-blocked from being at zero orbital angular momentum. For this proof-of-principle study, it is assumed that the two and three-body interactions between nucleons and $\Lambda$-baryons are the same. Figure~\ref{fig:infvol} and  Table \ref{tab:inf} summarise the results of this work as well as previous EFT matching calculations from Refs.~\cite{Barnea:2013uqa,Eliyahu:2019nkz,Detmold:2021oro}.
The $A=4$ EFT calculations are post-dictions of the infinite-volume-extrapolated LQCD results of Ref.~\cite{Beane:2012vq}, while the $A\in\{5,6\}$ results are predictions that could be tested by future LQCD calculations.

\section{Summary and Outlook}
\label{sec:conclusion}

In this work, differentiable programming and a generalised eigenvalue problem have been used to optimise the ground state wavefunctions of nuclei in FVEFT matched to LQCD binding energies. Using sets of correlated Gaussian wavefunctions representing $A$-nucleon states for $2\le A\le 6$ in both finite and infinite volumes, it was shown that this approach provides a considerably more efficient representation of these states than that obtained in previous work using the stochastic variational method and is able to scale to larger system size for fixed computational resources. 

Ongoing work to extend this approach by coupling spatial and spin wavefunctions used for the nuclear states will allow more physical systems to be addressed including $p$-shell nuclei and hypernuclei. Since these approaches can also provide accurate representations of finite-volume excited states, a more detailed matching to the low energy excitation spectra of two-nucleon systems in LQCD, for example those presented in Ref.~\cite{Amarasinghe:2021lqa}, will also allow more precise constraints on the LECs of the nuclear EFT, including those that occur at next-to- and next-to-next-to-leading order in the EFT power-counting. 

Pionless EFT is particularly powerful at the large quark masses used in existing LQCD calculations of nuclei. However, as the masses used in such calculations become closer to the physical light-quark masses, extensions of the finite-volume matching approach presented here to chiral EFTs that include explicit pionic degrees of freedom will likely be important. Alternative finite-volume many-body methods such as quantum Monte-Carlo \cite{Carlson:2014vla,Gandolfi:2017arm} and nuclear lattice EFT \cite{Epelbaum:2010xt,Lahde:2019npb} are promising approaches.

\acknowledgements

We thank Betzalel Bazak and Fernando Romero-L\'opez for insightful discussions and comments. This work is supported by the National Science Foundation under Cooperative Agreement PHY-2019786 (The NSF AI Institute for Artificial Intelligence and Fundamental Interactions, http://iaifi.org/). WD and PES are supported in part by the U.S.~Department of Energy, Office of Science, Office of Nuclear Physics under grant Contract Number DE-SC0011090. WD is also supported by the SciDAC4 award DE-SC0018121. PES is additionally supported by the National Science Foundation under EAGER grant 2035015, and by the U.S. DOE Early Career Award DE-SC0021006.
The authors acknowledge the MIT SuperCloud and Lincoln Laboratory Supercomputing Center for providing NVidia V100 GPU  resources that have contributed to the research results reported herein \cite{reuther2018interactive}. 

\clearpage

\onecolumngrid

\appendix

\section{Integral definitions and automatic differentiation}
\label{sec:appDP}

\def\bcut{\tilde{b}}
\def\qcut{\tilde{q}}

This section presents explicit analytic formulae for the matrix elements required to compute variational energy bounds with Gaussian wavefunctions, and provides details of the computation of the derivatives of those matrix elements as required for the DP optimisation procedure.

The normalisation matrix $\mathbb{N}$ (defined in Eq.~\eqref{eq:HNint}), with elements labelled by the $i$-th and the $j$-th terms of wavefunctions, can be computed as
\begin{align}
    [\mathbb{N}]_{ij} \equiv &\int \Psi_L^\text{sym}(A_i,B_i,{\bf d}_i;{\bf{x}})^* \Psi_L^\text{sym}(A_j,B_j,{\bf d}_j;{\bf{x}}) d{\bf{x}}\nonumber\\
    = &\sum_{\mathcal{P},\mathcal{P}'} \prod_{\alpha\in\{x,y,z\}}
    \sqrt{\frac{(2\pi)^{n}}{\text{Det}[C^{(\alpha)}_{i\mathcal{P};j\mathcal{P}'}]}}
    \sum_{{\bf b}^{(\alpha)}}^{|{\bf b}^{(\alpha)}|\le \bcut}
   \text{exp}\Bigg[
    -\frac{1}{2} \Omega^{(\alpha)}_{i\mathcal{P};j\mathcal{P}'}
    \Bigg],
    \label{eq:Nij}
\end{align}
where ${\cal P}$ and  ${\cal P}{'}$ denote permutations over the $n$ bodies in each wavefunction and  $\Omega^{(\alpha)}_{i\mathcal{P};j\mathcal{P}'}$ and 
$C^{(\alpha)}_{i\mathcal{P};j\mathcal{P}'}$ are quantities defined in Ref.~\cite{Detmold:2021oro} that depend on the parameters of the wavefunctions. The finite-volume sum is controlled by the integer cutoff $\bcut$.

The matrix representation of the Hamiltonian $\mathbb{H}$ (Eq.~\eqref{eq:Hint}) can be broken up into three parts as
\begin{align}
    \mathbb{H}  = & {}\mathbb{K}+\mathbb{V}_2+\mathbb{V}_3,
\end{align}
where $\mathbb{K}$, $\mathbb{V}_2$, and $\mathbb{V}_3$ denote the kinetic energy, the two-body potential, and the three-body potential, respectively.
These are given by
\begin{align}
    [\mathbb{V}_2]_{ij} &\equiv C \sum_{a<b}^{n} \int \Psi_L^\text{sym}(A_i,B_i,{\bf d}_i;{\bf{x}})^* g_\Lambda({\bf x}_{a}-{\bf x}_{b},L)\Psi_L^\text{sym}(A_j,B_j,{\bf d}_j;{\bf{x}}) d{\bf{x}} \nonumber \\
    & {}= C \frac{\Lambda^{3}}{8 \pi^{3 / 2}} \sum_{\mathcal{P},\mathcal{P}'}\sum_{a<b}^{n}\prod_{\alpha\in\{x,y,z\}} \sqrt{\frac{(2\pi)^{n}}{\text{Det}[C^{(\alpha)}_{i\mathcal{P};j\mathcal{P}'}]}}\sqrt{\frac{\widetilde{C}^{(\alpha)}_{i\mathcal{P};j\mathcal{P}'}}{\widetilde{C}^{(\alpha)}_{i\mathcal{P};j\mathcal{P}'}+2\rho}}
    \sum_{{\bf b}^{(\alpha)}}^{|{\bf b}^{(\alpha)}|\le \bcut}
    \text{exp}\Bigg[ 
    -\frac{1}{2} \Omega^{(\alpha)}_{i\mathcal{P};j\mathcal{P}'}\Bigg]
    \nonumber \\
    & \qquad \times \sum_{q^{(\alpha)}=-\qcut}^{\qcut}
    \text{exp}\Bigg[- \frac{\rho\,\widetilde{C}^{(\alpha)}_{i\mathcal{P};j\mathcal{P}'} }{\widetilde{C}^{(\alpha)}_{i\mathcal{P};j\mathcal{P}'}+2\rho}\left([(C^{(\alpha)}_{i\mathcal{P};j\mathcal{P}'})^{-1} \cdot \mathbf{\Xi}^{(\alpha)}]_a - [(C^{(\alpha)}_{i\mathcal{P};j\mathcal{P}'})^{-1} \cdot \mathbf{\Xi}^{(\alpha)}]_b - Lq^{(\alpha)}\right)^2
    \Bigg],\label{eq:v2}
\end{align}
where $C= \frac{1}{2}n(n-1)C_0 +2(S_h(S_h+1)-\frac{3}{4}n)C_1$ for an $n$-body  nucleus of spin  $S_h$,
\begin{align}
    [\mathbb{V}_3]_{ij} &\equiv D_0\sum_{a\ne b\ne c}^{\rm cyc} \int \Psi_L^\text{sym}(A_i,B_i,{\bf d}_i;{\bf{x}})^* g_\Lambda({\bf x}_{a}-{\bf x}_{b},L)g_\Lambda({\bf x}_{b}-{\bf x}_{c},L)\Psi_L^\text{sym}(A_j,B_j,{\bf d}_j;{\bf{x}}) d{\bf{x}} \nonumber \\
     =&{} D_0\left(\frac{\Lambda^{3}}{8 \pi^{3 / 2}}\right)^2 \sum_{\mathcal{P},\mathcal{P}'}\sum_{a\ne b\ne c}^{\rm cyc} \prod_{\alpha\in\{x,y,z\}}
    \sqrt{\frac{(2\pi)^{n}}{\text{Det}[{\widehat{C}}^{(\alpha)}_{i\mathcal{P};j\mathcal{P}'}]}} 
\text{exp}\left[-\frac{1}{2}\left(
    {\bf d}^{(\alpha)}_{i\mathcal{P}} \cdot B^{(\alpha)}_{i\mathcal{P}} \cdot {\bf d}^{(\alpha)}_{i\mathcal{P}} 
    + {\bf d}^{(\alpha)}_{j\mathcal{P}'} \cdot B^{(\alpha)}_{j\mathcal{P}'} \cdot {\bf d}^{(\alpha)}_{j\mathcal{P}'}
    \right)\right] 
    \nonumber \\
    & \qquad
\times \sum_{{\bf b}^{(\alpha)}}^{|{\bf b}^{(\alpha)}|\le \bcut}
\text{exp}\left[
-\frac{1}{2} \left(
(L{\bf b}^{(\alpha)})\cdot (A^{(\alpha)}_{i\mathcal{P}}+B^{(\alpha)}_{i\mathcal{P}}) \cdot (L {\bf b}^{(\alpha)})
+   2 {\bf d}^{(\alpha)}_{i\mathcal{P}} \cdot B^{(\alpha)}_{i\mathcal{P}} \cdot (L {\bf b}^{(\alpha)}) 
     -\mathbf{\Xi}^{(\alpha)} \cdot 
     [{\widehat{C}}^{(\alpha)}_{i\mathcal{P};j\mathcal{P}'}]^{-1}
     \cdot \mathbf{\Xi}^{(\alpha)}
     \right)\right]
     \nonumber \\
     & \qquad
     \times
     \sum_{q^{(\alpha)}=-\qcut}^{\qcut}
    \text{exp}\Bigg[
    -\frac{L^2}{r_0^2} q^{(\alpha)2}
    +\frac{q^{(\alpha)2}L^2}{2r_0^4}
    \mathfrak{P}_v^{[a, b]} \cdot 
    [{\widehat{C}}^{(\alpha)}_{i\mathcal{P};j\mathcal{P}'}]^{-1}
    \cdot \mathfrak{P}_v^{[a, b]}
    + \frac{q^{(\alpha)}L}{r_0^2} \mathbf{\Xi}^{(\alpha)} \cdot 
    [{\widehat{C}}^{(\alpha)}_{i\mathcal{P};j\mathcal{P}'}]^{-1} 
    \cdot \mathfrak{P}_v^{[a, b]}
    \Bigg]
    \nonumber\\
    & \qquad
    \times
     \sum_{t^{(\alpha)}=-\qcut}^{\qcut}
    \text{exp}\Bigg[
    -\frac{L^2}{r_0^2} t^{(\alpha)2}
    +\frac{t^{(\alpha)2}L^2}{2r_0^4}
    \mathfrak{P}_v^{[b, c]} \cdot 
    [{\widehat{C}}^{(\alpha)}_{i\mathcal{P};j\mathcal{P}'}]^{-1}
    \cdot \mathfrak{P}_v^{[b, c]}
    + \frac{t^{(\alpha)}L}{r_0^2} \mathbf{\Xi}^{(\alpha)} \cdot 
    [{\widehat{C}}^{(\alpha)}_{i\mathcal{P};j\mathcal{P}'}]^{-1} 
    \cdot \mathfrak{P}_v^{[b, c]}
    \Bigg] 
    \nonumber\\& \hspace{22mm}
    \times \text{exp}\left[\frac{t^{(\alpha)}q^{(\alpha)}L^2}{r_0^4}\mathfrak{P}_v^{[b, c]}\cdot [{\widehat{C}}^{(\alpha)}_{i\mathcal{P};j\mathcal{P}'}]^{-1}
    \cdot\mathfrak{P}_v^{[a, b]})\right]
    \,,\label{eq:v3}
\end{align}

\begin{align}
    [\mathbb{K}]_{ij} &\equiv -\frac{1}{2M_N} \sum_{a=1}^n \int \Psi_L^\text{sym}(A_i,B_i,{\bf d}_i;{\bf{x}})^*  \nabla_a^2\Psi_L^\text{sym}(A_j,B_j,{\bf d}_j;{\bf{x}}) d{\bf{x}} \nonumber \\
    & = 
   \frac{1}{2M_N}  \sum_{\mathcal{P},\mathcal{P}'}
     \sum_{\alpha\in\{x,y,z\}} \sqrt{\frac{(2\pi)^{n}}{\text{Det}[C^{(\alpha)}_{i\mathcal{P};j\mathcal{P}'}]}}
      \sum_{{\bf b}^{(\alpha)}}^{|{\bf b}^{(\alpha)}|\le \bcut}
      \Theta^{(\alpha)}_{i\mathcal{P};j\mathcal{P}'}
      \text{exp}\Bigg[
    -\frac{1}{2} \Omega^{(\alpha)}_{i\mathcal{P};j\mathcal{P}'}
    \Bigg]
   \nonumber  \\ & \hspace{6cm}
    \times\prod_{\beta\in\{x,y,z\}}^{\beta\ne\alpha}
    \sqrt{\frac{(2\pi)^{n}}{\text{Det}[C^{(\beta)}_{i\mathcal{P};j\mathcal{P}'}]}}
      \sum_{{\bf b}^{(\beta)}}^{|{\bf b}^{(\beta)}|\le \bcut}
      \text{exp}\Bigg[
    -\frac{1}{2} \Omega^{(\beta)}_{i\mathcal{P};j\mathcal{P}'}\Bigg],
        \label{eq:Kij}
\end{align}
where $\hbar=1$ is used, $\sum_{a\ne b\ne c}^{\rm cyc}$ indicates a sum over cyclic permutations of  particles $a$, $b$ and $c$, and the integer cutoff $\qcut$ governs finite-volume effects in the interaction terms.
The (wavefunction-parameter dependent) quantities
$\mathbf{\Xi}^{(\alpha)}_{i\mathcal{P};j\mathcal{P}'}$, $\widetilde{C}^{(\alpha)}_{i\mathcal{P};j\mathcal{P}'}$, ${\widehat{C}}^{(\alpha)}_{i\mathcal{P};j\mathcal{P}'}$ and $\Theta^{(\alpha)}_{i\mathcal{P};j\mathcal{P}'}$, as well as the projection operators $\mathfrak{P}_v^{[a, b]}$ and $\mathfrak{P}_m^{[a, b]}$, are defined in Ref.~\cite{Detmold:2021oro}.

The variational function $\mathcal{E}$ to be minimised (Eq.~\eqref{eq:variationalGS}) can be represented in terms of these matrices as
\begin{equation}
    \mathcal{E}\left[\Psi^{(N)}_h(\theta)\right] = \dfrac{\mathbf{c} \cdot ( \mathbb{K}+\mathbb{V}_2+\mathbb{V}_3 ) \cdot \mathbf{c}}{\mathbf{c} \cdot \mathbb{N} \cdot \mathbf{c}},
    \label{eq:Efunc}
\end{equation}
where ${\bf c}=(c_1,\ldots,c_N)^T$ collects the numerical coefficients ($c_i$ of Eq.~\eqref{eq:Efunc}) parameterising $\Psi^{(N)}_h(\theta)$ as a linear combination of the Gaussian wavefunction terms $\Psi_L^\text{sym}(A_i,B_i,{\bf d}_i;{\bf{x}})$, and $\theta=\{\{A_i,B_i,{\bf d}_i,c_i\}$, $i\in\{1,\ldots N\}\}$.
Storing the computational graph for Eq.~\eqref{eq:Efunc} and its gradients with respect to the parameters $\theta$ requires a large amount of memory.
To reduce the memory usage, the chain rule is applied manually to compute the gradient of $\mathcal{E}$ as
\begin{equation}
    \nabla_\theta \mathcal{E} = -\dfrac{\mathbf{c} \cdot (\mathbb{K}+\mathbb{V}_2+\mathbb{V}_3) \cdot \mathbf{c}}{(\mathbf{c} \cdot \mathbb{N} \cdot \mathbf{c})^2} \nabla_\theta (\mathbf{c} \cdot \mathbb{N} \cdot \mathbf{c}) + \dfrac{1}{\mathbf{c} \cdot \mathbb{N} \cdot \mathbf{c}}(\nabla_\theta(\mathbf{c} \cdot \mathbb{K} \cdot \mathbf{c}) + \nabla_\theta(\mathbf{c} \cdot \mathbb{V}_2 \cdot \mathbf{c}) + \nabla_\theta(\mathbf{c} \cdot \mathbb{V}_3 \cdot \mathbf{c})).
\end{equation}
The computation of $\nabla_\theta (\mathbf{c} \cdot \mathbb{X} \cdot \mathbf{c})$ for $\mathbb{X}\in\{\mathbb{N}, \mathbb{K}, \mathbb{V}_2,\mathbb{V}_3\}$ can be further broken up into a sum involving the gradient of each matrix element
\begin{equation}
    \nabla_\theta (\mathbf{c} \cdot \mathbb{X} \cdot \mathbf{c}) = \sum_{i,j}\left( \nabla_\theta (c_i c_j) [\mathbb{X}]_{ij}  + c_i c_j \nabla_\theta [\mathbb{X}]_{ij}\right).
\end{equation}
Due to the permutation symmetry in this system, there are only $n!n(n-1)$ independent terms in the summation over permutations ${\cal P}$ and ${\cal P}{'}$ in $\mathbb{V}_2$ (Eq.~\eqref{eq:v2}) and  $n!n(n-1)(n-2)$ terms in $\mathbb{V}_3$  (Eq.~\eqref{eq:v3}). Their gradients can be written as a sum of gradients on each independent term whose computational graph is discarded after its gradient is computed.

\section{Numerical implementation details}
\label{app:op}

A key component of the calculations presented here is the evaluation of the matrix elements $\mathbb{N}_{ij}$ and $\mathbb{H}_{ij}$ that enter both the DP and GEVP blocks. 
The numerical accuracy of these matrix elements is controlled by the integer cutoffs $\bcut$ and $\qcut$ used in the summations in Eqs.~\eqref{eq:Nij}--\eqref{eq:Kij}; for small values of these cutoffs, numerical instabilities appear with $\mathbb{N}$ potentially becoming non-positive-definite. Since the goal of the DP block is simply to produce trial wavefunctions, the accuracy criteria on matrix elements in the DP block is somewhat milder than in the GEVP block where a rigorous energy bound is sought. Consequently, 
$\tilde{b}=15$ and $\tilde{q}=6$ are used during  automatic differentiation and $\tilde{b}=30$ and $\tilde{q}=12$ are chosen for the solution of the GEVP.
These values avoid numerical stability issues but allow evaluation of the matrix elements for $N\in \{2,3\}$-body systems at finite volume.  For the optimisation and the solution of GEVP of four-body system binding energies $\tilde{b}=8$ and $\tilde{q}=3$ are chosen due to computational limitations.

The DP process depends on the initialisation of the wavefunction parameters $\theta$. As in Ref.~\cite{PhysRevA.87.063609}, the matrices $A$ and $B$ are generated from single-particle Gaussian widths $d_a$ and two-body Gaussian widths $d_{ab}$.   The particle displacement vectors $\bf{d}$ have components $d_i$ and the weights of each wavefunction are written as $c_j=\tan{\hat{c}_j}$ to ensure both positive and negative values are accessed. The parameters $d_a$, $d_{ab}$, $d_i$ and $\hat{c}_j$ are drawn from a normal distribution $N(1,0.01)$, which in practice leads to stable results.

In the optimisation step in the DP block, a self-adaptive gradient descent method with a stepping clip is applied. For each step, the learning rate is increased by $20\%$ if the energy decreases but is decreased by $60\%$ if the energy increases. Steps in which the energy increases are rejected. In addition, a maximum allowed step size is implemented for each parameter. The step in parameter $\theta_i\in\theta$ is $-\eta \partial_{\theta_i} \mathcal{E}$ if its absolute value is smaller than $f(\eta)$. Otherwise, the change in $\theta_i$ is $-\textrm{sgn}(\partial_{\theta_i} \mathcal{E}) f(\eta)$, where
\begin{equation}
    f(\eta) = \left\{ 
    \begin{aligned}
        10^{-2},\ &\eta >0.2\\
        10^{-3},\ &0.001 < \eta \le 0.2\\
        10^{-4},\ &\eta \le 0.001
    \end{aligned}
    \right..
\end{equation}

Since the uncertainties of the LQCD results increase with $A$, the precision necessary in the variational optimisation for optimisation uncertainties to be sub-dominant relative to the LQCD uncertainties, or to the uncertainties propagated from the matching of the two and three-body LECs, decreases with $A$.
For two and three-body systems, the training process is iterated until the relative change of the energy over the last 10 iterations is less than $10^{-5}$. Under this condition, the relative differences between the upper bounds obtained using different seeds and the same set of LECs are less than $0.5\%$ for all of the results presented in Sections \ref{sec:3b}--\ref{sec:3d}. 
For the four-body system, the relative changes in the last 10 steps of optimisation are less than $10^{-4}$ and the variations between initialisations are  less than $1\%$.
For the infinite volume calculations, the same convergence bounds hold for the two, three, and four-body systems. For five and six-body systems, the relative changes over the last ten steps are less than $10^{-3}$, and the relative differences between results obtained with different initialisations are less than $2\%$.

\bibliography{reference}
\end{document}